# CH$_4$·F$^-$ revisited: Full-dimensional *ab initio* potential energy surface and variational vibrational states

Dóra Papp,[1,*] Viktor Tajti,[1] Gustavo Avila,[2,*] Edit Mátyus,[2,*] and Gábor Czakó[1,*]

[1] MTA-SZTE Lendület Computational Reaction Dynamics Research Group, Interdisciplinary Excellence Centre and Department of Physical Chemistry and Materials Science, Institute of Chemistry, University of Szeged, Rerrich Béla tér 1, H-6720 Szeged, Hungary

[2] ELTE, Eötvös Loránd University, Institute of Chemistry, Pázmány Péter sétány 1/A, 1117 Budapest, Hungary

[*] E-mail: dorapapp@chem.u-szeged.hu (D. P.), Gustavo_Avila@telefonica.net (G. A.), edit.matyus@ttk.elte.hu (E. M.), gczako@chem.u-szeged.hu (G. C.)

## Abstract

The automated development of a new *ab initio* full-dimensional potential energy surface (PES) is reported for the CH$_4$·F$^-$ complex using the RoboSurfer program package. The new potential provides a near-spectroscopic quality description over a broad configuration range including the methane-ion dissociation, as well as isolated methane vibrations. In particular, it improves upon the earlier [Czakó, Braams, Bowman (2008)] PES over intermediate methane-fluoride distances. Full-dimensional (12D) variational vibrational computations using the new PES and the GENIUSH-Smolyak algorithm show that tunneling splittings larger than 0.1 cm$^{-1}$ appear below the top of the interconversion barrier of the four equivalent minima of the complex.

# I. INTRODUCTION

Three years ago, the first, full-dimensional vibrational computation that converged tunneling splittings of the 12D $CH_4 \cdot F^-$ complex was reported on the occasion of Attila Császár's 60th birthday [1]. Further progress was hindered due to an incorrect behaviour of the potential energy surface over intermediate ion-molecule separations. This year, Péter Szalay, an electronic theory expert and good colleague from the same department, turns sixty. On this occasion, we are happy to report the development of a new and well-behaved *ab initio* potential energy surface of spectroscopic quality over a broad range of nuclear configurations. This development opens the route for testing and further developing the vibrational methodology towards the spectroscopy of the methane fragment in this complex, and possibly, also for other molecular systems and complexes.

Molecular complexes are the simplest systems in which molecular interactions occur. Molecular interactions play a central role behind the functionality of biological systems and properties of materials. Molecular interactions can be probed by spectroscopic techniques. Direct probing of the intermolecular dynamics in the low-frequency range of the infrared and Raman spectra is often challenging for practical reasons and also in terms of interpretation of the experimental results. Observation and interpretation of shifts and splittings of the characteristic frequencies in the infrared range, *e.g.*, C–H or O–H stretching of the intramolecular monomer modes is more common than direct probing of the intermolecular dynamics in the far-infrared range. Rich experimental information on the intramolecular vibrations of molecular complexes is available not only for the fundamental modes but also for the overtones of the constituent molecules [2].

At the same time, (variational) (ro)vibrational results for the intramolecular (often predissociative) modes are rarely available. Important, recent exceptions are due to Felker, Bačić, and their co-workers who used and developed contraction techniques [3, 4, 5] and computed (ro)vibrational states of the water molecule in complexes with (rigid) benzene [6], hydrogen chloride [7, 8], and carbon monoxide [6, 9].

Vibrational computations of high(er)-dimensional molecular systems with floppy modes have been challenging due to the dimensionality and poor separability of the problem, and in general, due to the exponential scale-up of exact quantum dynamics methodologies with the number of the coupled vibrational degrees of freedom. Over the past decade, development is observed in several important directions, (a) coordinate representation and the kinetic energy operator [10, 11, 12, 13, 14, 15, 16, 17]; (b) contraction techniques [3, 4, 5, 6, 7], (c) grid-pruning techniques [1, 18, 19, 20, 21, 22]; (d) collocation [23, 24, 25]; and (e) PES representation for high-dimensional quantum dynamics [26, 27, 28]; and (f) highly parallel computation of millions of states [29, 30, 31]. Methodological progress makes it possible to attenuate the exponential scale-up, including systems with large-amplitude motions [1, 21, 32]. Variational vibrational computations start from the bottom of the vibrational energy spectrum, *i.e.*, first, the intermolecular fingerprint range is accessed.

To reach and more importantly, to specifically target the intramolecular modes in the complex, further developments of the vibrational methodology is needed. The fluoride-methane complex is an excellent work horse for these developments. In comparison with methane-argon [1, 32], the strong and anisotropic interaction with the fluoride introduces substantial distortions of the methane fragment and the interaction couples several methane rovibrational states.

The fluoride-methane complex has attracted some interest also due to the possibility to observe 'heavy-atom' tunneling. Some controversy regarding the size of the tunneling splittings from various models and computations [1, 33, 34, 35] was resolved with the direct computation and convergence of the tunneling splittings of the fully coupled 12-dimensional problem for the low-energy range [1]. Experimental results are available for the higher-energy (predissociative) methane intramolecular vibrational range [36, 37, 38, 39] that motivates further development of the vibrational methodology.

Regarding earlier theoretical work, Czakó, Braams, and Bowman developed the first, full-dimensional *ab initio* potential energy surface (PES) in 2008 [33] (henceforth labelled as CBB08). On CBB08, the complex is bound by an $D_e = 2434$ cm$^{-1}$ electronic dissociation energy with an equilibrium structure in which the fluoride is closest to one of the hydrogens of the methane fragment. The complex has four equivalent minima with $C_{3v}$ point-group symmetry. Ref. [34] estimated the interconversion barrier connecting the equivalent wells to be ∼1270 cm$^{-1}$.

Regarding earlier vibrational studies, Czakó, Braams, and Bowman carried out full-dimensional vibrational computations using the MULTIMODE program and normal coordinates corresponding to one of the PES minima. The zero-point energy-corrected dissociation energy was obtained to be $D_0 = 2316$ cm$^{-1}$. This initial work was followed by 12D multi-configuration time-dependent Hartree (MCTDH) computations [34], 3D GENIUSH computations [35], and 12D GENIUSH-Smolyak [21] computations [1]. Ref. [1] provided the first rigorous assessment about possible 'heavy-atom' tunneling and the corresponding splittings of the vibrational frequencies. Ref. [1] has shown that already below the top of the 'isomerisation' barrier tunneling splittings between 0.1–1 cm$^{-1}$ appear. In Ref. [1], it was also noted however that the CBB08 PES has an incorrect behaviour for intermediate separation of the CH$_4$ and the F$^-$ that hindered further progress with the vibrational computations.

The present work reports the development of a new PES that describes correctly the dissociation of the CH$_4$ and the F$^-$ fragments of this strongly anisotropic complex. The details of the PES development and the properties of the PES are described in Section II. In Section III the results obtained from full-dimensional variational vibrational computations on the new PES are presented. The paper ends with summary and conclusions in Section IV.

## II. CONSTRUCTION OF THE POTENTIAL ENERGY SURFACE

### A. Initial PES development

As a first step, the global minimum (featuring one H-bond) of the PES, having $C_{3v}$ point-group symmetry, is determined by performing geometry optimization and harmonic frequency computations at the explicitly-correlated CCSD(T)-F12b/aug-cc-pVTZ[40,41] level of theory, which is selected for computing the energy points of the PES. For *ab initio* computations we use the MOLPRO program package [42].

To cover the spectroscopically interesting region of the full-dimensional PES of the CH$_4$·F$^-$ complex, we scatter the F$^-$ ion isotropically around the methane molecule in the range of [2.5, 50.0] bohr while the Cartesian coordinates of the atoms of the CCSD(T)-F12b/aug-cc-pVTZ optimized methane were randomly displaced by values in the range of [−0.94, +0.94] bohr. To further improve the PES, we also carry out random displacements in the normal coordinates of methane while scattering the F$^-$ ion isotropically, within [2.5, 30.0] bohr, concentrating especially on the [4.0, 8.0] bohr interval, around the displaced methane molecule. The normal coordinates are randomized in two ways: (1) one of the $t_2$-symmetry and one of the $e$-symmetry deformation, the symmetric stretching, and two of the $t_2$-symmetry asymmetric stretching normal coordinates are separately given a uniformly distributed random value generated in the intervals specified in **Table I**, and (2) all the above selected five normal coordinates are given random values simultaneously using also the intervals of **Table I** (in the case of the *e*-symmetry deformation and the symmetric stretching coordinate we use the interval given in parenthesis to reduce the overall displacements). The random normal coordinates correspond to Cartesian displacements from the equilibrium structure of methane on the current version of the PES under development when the F⁻ ion is placed far (57 bohr).

**Table I.** Intervals of random displacements of the normal coordinates ($u^{1/2}$Å) of methane to generate the initial geometry set for PES development. In parantheses the intervals applied in simultaneous displacements are shown (see text for details).

| Normal coordinate (symmetry) | Interval |
|:---:|:---:|
| deformation ($t_2$) | −1.2 – 0.8 |
| deformation ($e$) | −3.0 – 1.2 (−2.0 – 1.2) |
| symmetric stretching ($a_1$) | − 0.6 – 3.0 (−0.6 – 2.0) |
| asymmetric stretching ($t_2$) | −0.4 – 0.4 and −0.6 – 0.6 |

As an additional improvement we add energy points to the PES from two-dimensional grids constructed by imitating CH-stretching motions (first dimension) of methane of the CCSD(T)-F12b/aug-cc-pVTZ global minimum geometry and placing the F⁻ ion collinearly with one C-H bond in the range of [4.5, 9.0] bohr (second dimension). Extremely high-energy points are sorted out from the generated geometry set. At this stage, we arrive at a PES consisting of 24 209 geometries and corresponding energies computed at the CCSD(T)-F12b/aug-cc-pVTZ level of theory.

The PES is fitted using the permutationally invariant polynomial method[43], based on polynomials of Morse-like variables of the $r_{ij}$ internuclear distances $y_{ij} = \exp(−r_{ij}/a)$ with $a = 3.0$ bohr. The full-dimensional analytical function, representing the PES, which is invariant under the permutation of like atoms, has a highest total order of the polyanomials of 7. We note that a 7$^{th}$-order polynomial of $y_{ij}$ can accurately describe the asymptotic region of the PES, even better than a single-parameter ($\sigma$) $−\sigma/r_{ij}^n$ function, as demonstrated for the CH$_4$·Ar complex, where $n = 6$, in Ref. [32]. The PES is determined using 9355 coefficients in a weighted least-squares fit on the energy points, with a weight of ($E_0/(E_0 + E))\times(E_1/(E_1 + E)$) with $E_0 = 0.05$ hartree and $E_1 = 0.5$ hartree applied on a given energy $E$ relative to the global minimum of the data set.

## B. Automated PES improvement with ROBOSURFER

The PES fitted on the energy points generated as detailed above has first been subjected to preliminary vibrational quantum dynamics computations. From these test computations it has turned out that although the PES describes excellently the dissociation of the $F^-$ ion from methane, the vibrations of methane, especially near the global minimum structure, are not described properly enough. This finding is somewhat unsurprising, as the $CH_4 \cdot F^-$ complex is much more strongly bound than the $CH_4 \cdot Ar$ adduct, for which the present authors have already successfully developed an accurate PES by only scattering the Ar atom around the randomly displaced methane molecule [32].

After realizing the deficiencies of our initial PES, we have set up the ROBOSURFER program system [44], developed in the Czakó group and so far used only for the automated improvement of reactive PESs constructed for quasi-classical trajectory (QCT) simulations [45,46,47,48], and have started to further augment the PES for the $CH_4 \cdot F^-$ complex. Briefly, ROBOSURFER, as shown schematically in **Figure 1**, follows a fully automated strategy of generating new geometries based on QCT computations on a current version of the PES under development, and thoroughly selecting the ones that improve most effectively the fitting: geometric and energetic similarity used to estimate the fitting error of the new geometries [44]. It also performs quantum chemical computations automatically, then adds new geometries to the preexisting PES iteratively, usually the ones with the largest fitting errors. ROBOSURFER also features a subprogram, called HOLEBUSTER, that eliminates artificial minima on the surface, as well as a rebuilding protocol, that can be highly beneficial if the reduction of the initial (or some intermediate) dataset is desired.

We start the automated development with a rebuilding procedure, by fitting approximately only the quarter of the full data set (*set A*), and let ROBOSURFER to "choose" from the remaining geometries. After doubling *set A* this way, we discard all the initial geometries of *set A*, and let the program to re-select among them. At this point, we have 12 237 geometries in the fitting set to start the actual development process.

Throughout the development of the PES with ROBOSURFER, we apply a minimal collision energy between $CH_4$ and $F^-$, only 100 cm$^{-1}$, to be able to run the quasi-classical dynamics calculations. In the first 15 iterations the $F^-$ is placed far (25 Å) and $CH_4$ is in its harmonic ground vibrational state, imitating the independently vibrating methane, and we only allow a 2000 maximum step size during the trajectories. The ZPE of the initial structure for the trajectory is set by standard normal-mode sampling [49], and the orientations of the fragments is random. The distance of the fragments is $\sqrt{x^2 + b^2}$, where $b$ is the so-called impact parameter. In our trajectories we apply $b = 0.0$ and $0.5$ bohr, and 40 trajectories are run in each ROBOSURFER iteration. For the next 104 iterations, we start the trajectories from the H-bonded global minimum structure ($x = 2.954$ Å, optimized on the current PES) with a maximal step number of 2000. We than increase the step number to 100 000, and the distance of the reactants to $x = 15$ Å, to imitate their collisions through an additional 38 iterations. The final PES dataset contains 13 065 geometries and the corresponding CCSD(T)-F12b/aug-cc-pVTZ energies.

### C. Properties of the potential energy surface

The final PES features 3.12 and 3.40 cm$^{-1}$ root-mean-squares (RMS) fitting errors for the 0−11 000 and the 11 000−22 000 cm$^{-1}$ energy intervals with respect to the global minimum, respectively. The $D_e$ dissociation energy of the H-bonded global minimum complex computed on the PES is 2518.26 cm$^{-1}$, while the *ab initio* $D_e$ calculated at the CCSD(T)-F12b/aug-cc-pVTZ level of theory is 2519.90 cm$^{-1}$, which comparison highlights the remarkable accuracy of the fit. An edge-bound $C_{2v}$ first-order saddle point (266$i$ cm$^{-1}$) with electronic barrier height of 1269 cm$^{-1}$ connects the equivalent face-bound $C_{3v}$ minima. This value is almost the same as the one (~1270 cm$^{-1}$) found on the CBB08 PES [33] in [34] (we note that in Ref. [1] the barrier height was incorrectly reported, the correct value is 1270 cm$^{-1}$).

In **Figure 2**, we plot various one-dimensional (1D) potential energy curves recorded during the automated development of the PES with ROBOSURFER. *Ab initio* energies determined at the CCSD(T)-F12b/aug-cc-pVTZ level of theory and energy values obtained on the PES are compared also in **Figure 2** along these potential curves. The geometries representing the different 1D cuts are generated as follows: (1) single C-H stretch: one C-H bond among the three C-H bonds not taking part in the H-bond of the global minimum structure is stretched, (2) triple C-H stretch: all C-H bonds, apart from the one forming the H-bond are stretched, (3) H-bond C-H stretch: the C-H bond participating in the H-bond is stretched, (4) all H atoms are stretched along the respective C-H axes of the CCSD(T)-F12b/aug-cc-pVTZ equilibrium geometry of the global minimum or that of methane, when the F$^-$ ion is placed far along the H-bond axis (asymptotic region). Of course the two single C-H stretch curves should be the same in the asymptotic region. From **Figure 2**, it is clear that ROBOSURFER had a major job to correct the single C-H stretches, except the one obtained near the global minimum. The triple and symmetric C-H stretching potential cuts near the global minimum also seem to improve greatly during the automated development, while needed only minor corrections in the asymptotic region. There are still some breakdowns in the PES at C-H distances at larger *r* values, but they are confirmed not to cause any problems during vibrational quantum dynamics computations.

In **Table II**, we show the variationally determined vibrational energy levels of isolated methane obtained on our PES before and after the automated development compared to the spectroscopically accurate reference results of Wang and Carrington [50]. After the automated procedure, the energy levels are in much better agreement with the reference values, highlighting the significant improvement accomplished by using ROBOSURFER on the methane vibrations. Moreover, the ROBOSURFER PES can be evaluated about 61 times faster (see **Table II**), because the PES version before the automated sampling required the use of quadruple precision to avoid unphysical numerical oscillations in the potential energy. The present observation that machine-learning sampling speeds up PES evaluation was also found in Ref. [51]. The results unambiguously show that the comprehensive approach of ROBOSURFER, so far only used to develop reactive PESs constructed for reaction dynamics simulations, can be an effective tool in making PES development for spectroscopic purposes easier and more accurate, as well.

The potential energy cut of the final PES related to the asymptotic behavior of the PES, *i.e.*, when the $F^-$ ion is dissociating from methane (in this particular case along the H-bond axis), which was already excellent before the automated development as well, is plotted in **Figure 3**. The first panel shows that the PES 1D cut fits accurately on the *ab initio* CCSD(T)-F12b/aug-cc-pVTZ energies near the minimum, while the second panel highlights the excellent description of the asymptotic region of the PES (which is cut at 30 bohr on the figure, but proven to be accurate up to *r*(C-F) = 50 Å). The maximum difference between the 1D curve of the PES and the *ab initio* data is 5.3 cm$^{-1}$, and the average deviation is consistent with the RMS fitting error of about 3 cm$^{-1}$.

The geometry of the global minimum along with its free structural parameters obtained on the PES compared to the CCSD(T)-F12b/aug-cc-pVTZ values, are shown in **Figure 4**. All bond lengths agree within less than 0.001 Å, and the angles obtained with the two methods are also the same within two decimal places, underlining the accuracy of the fit.

**Table II**. Vibrational energy levels of the methane molecule in the complex at large methane-fluoride distance computed with the PESs developed in this work and the GENIUSH-Smolyak approach [1,21] in comparison with isolated methane values [50,52]. The energies, in cm$^{-1}$, are referenced to the zero-point vibrational energy (ZPVE).

| | CH$_4$ in CH$_4$·F$^-$ at fixed $R$ [a] | | | Isolated CH$_4$ | | |
|---|---|---|---|---|---|---|
| | Without ROBOSURFER | ROBOSURFER PES | | Variational [50] | Expt. [52] | |
| $t$/s [b] | 61 000 | 1000 | 1000 | | | |
| $R$/bohr | 15 | 15 | 70 | | | |
| [ZPVE] | [12211.67] | [12187.13] | [12211.02] | [9651.29] | | |
| 1 | 1312.38 | 1311.16 | 1311.16 | 1310.47 | 1310.76 | (0001)F$_2$ |
| 2 | 1312.38 | 1311.16 | 1311.16 | 1310.47 | 1310.76 | |
| 3 | 1312.38 | 1311.16 | 1311.16 | 1310.47 | 1310.76 | |
| 4 | 1535.61 | 1533.65 | 1534.81 | 1533.47 | 1533.33 | (0100)E |
| 5 | 1535.61 | 1533.65 | 1534.81 | 1533.47 | 1533.33 | |
| 6 | 2589.26 | 2586.90 | 2587.28 | 2586.99 | 2587.04 | (0002)A$_1$ |
| 7 | 2617.44 | 2614.68 | 2615.21 | 2613.83 | 2614.26 | (0002)F$_2$ |
| 8 | 2617.44 | 2614.68 | 2615.21 | 2613.83 | 2614.26 | |
| 9 | 2617.44 | 2614.68 | 2615.21 | 2613.83 | 2614.26 | |
| 10 | 2627.42 | 2624.73 | 2623.88 | 2624.44 | 2624.62 | (0002)E |
| 11 | 2627.42 | 2624.73 | 2623.88 | 2624.44 | 2624.62 | |
| 12 | 2837.42 | 2829.89 | 2831.55 | 2830.20 | 2830.32 | (0101)F$_2$ |
| 13 | 2837.42 | 2829.89 | 2831.55 | 2830.20 | 2830.32 | |
| 14 | 2837.42 | 2829.89 | 2831.55 | 2830.20 | 2830.32 | |
| 15 | 2850.77 | 2847.48 | 2848.93 | 2845.96 | 2846.07 | (0101)F$_1$ |
| 16 | 2850.77 | 2847.48 | 2848.93 | 2845.96 | 2846.07 | |
| 17 | 2850.77 | 2847.48 | 2848.93 | 2845.96 | 2846.07 | |
| 18 | 2914.29 | 2907.87 | 2907.87 | 2917.18 | 2916.48 | (1000)A$_1$ |
| 19 | 3015.00 | 3014.62 | 3015.46 | 3019.43 | 3019.49 | (0010)F$_2$ |
| 20 | 3015.00 | 3014.62 | 3015.46 | 3019.43 | 3019.49 | |
| 21 | 3015.00 | 3014.62 | 3015.46 | 3019.43 | 3019.49 | |
| 22 | 3069.02 | 3065.57 | 3067.31 | 3063.77 | 3063.65 | (0200)A$_1$ |
| 23 | 3070.73 | 3066.06 | 3068.12 | 3065.30 | 3065.14 | (0200)E |
| 24 | 3070.73 | 3066.06 | 3068.12 | 3065.30 | 3065.14 | |
| 25 | 3870.35 | 3869.75 | 3871.23 | 3870.57 | 3870.49 | (0003)F$_2$ |
| 26 | 3870.35 | 3869.75 | 3871.23 | 3870.57 | 3870.49 | |
| 27 | 3870.35 | 3869.75 | 3871.23 | 3870.57 | 3870.49 | |
| 28 | 3912.70 | 3909.34 | 3910.45 | 3908.82 | 3909.20 | (0003)A$_1$ |
| 29 | 3925.43 | 3920.17 | 3920.66 | 3920.24 | 3920.51 | (0003)F$_1$ |
| 30 | 3925.43 | 3920.17 | 3920.66 | 3920.24 | 3920.51 | |
| 31 | 3925.43 | 3920.17 | 3920.66 | 3920.24 | 3920.51 | |
| 32 | 3933.42 | 3930.71 | 3929.49 | 3930.99 | 3930.92 | (0003)F$_2$ |
| 33 | 3933.42 | 3930.71 | 3929.49 | 3930.99 | 3930.92 | |
| 34 | 3933.42 | 3930.71 | 3929.49 | 3930.99 | 3930.92 | |
| 35 | 4112.08 | 4099.61 | 4101.75 | 4101.46 | 4101.39 | (0102)E |
| 36 | 4112.08 | 4099.61 | 4101.75 | 4101.46 | 4101.39 | |
| 37 | 4140.96 | 4127.96 | 4130.87 | 4128.69 | 4128.76 | (0102)F$_1$ |
| 38 | 4140.98 | 4127.96 | 4130.87 | 4128.69 | 4128.76 | |
| 39 | 4140.98 | 4127.96 | 4130.87 | 4128.69 | 4128.76 | |
| 40 | 4140.98 | 4131.47 | 4132.31 | 4132.60 | 4132.86 | (0102)A$_1$ |
| 41 | 4152.37 | 4144.10 | 4146.53 | 4142.67 | 4142.87 | (0102)F$_2$ |
| 42 | 4152.37 | 4144.10 | 4146.53 | 4142.67 | 4142.87 | |
| 43 | 4152.37 | 4144.10 | 4146.53 | 4142.68 | 4142.87 | |
| 44 | 4157.28 | 4151.76 | 4152.72 | 4151.21 | 4151.21 | (0102)E |
| 45 | 4157.28 | 4151.76 | 4152.72 | 4151.21 | 4151.21 | |
| 46 | 4168.02 | 4163.23 | 4164.23 | 4161.89 | 4161.85 | (0102)A$_2$ |

[a] The potential energy was averaged over the four equivalent fluoride positions (hence the methane degeneracies are obtained exactly). For the methane basis, the $B = 10$ pruning condition was used (Sec. III).

[b] Computing time (for 1 processor core) for the evaluation of the PES over the quadrature points used for the variational vibrational computation.

## III. VARIATIONAL VIBRATIONAL COMPUTATIONS

### A. Methodological details

We aim to compute several dozens of vibrational states of the vibrational Schrödinger equation,

$$(\hat{T}^V + V)\Psi = E\Psi, \tag{1}$$

where $V$ is the PES developed during the first part of this work. We intend to include the PES in the vibrational computation without further expansions or approximations.

The vibrational kinetic energy operator in terms of general vibrational coordinates can be written in various forms [11]. The Podolsky form has been found to be useful [11, 53, 54, 55] in discrete variable representation (DVR) [56] computations,

$$\hat{T}^V_{\text{Pod}} = -\frac{1}{2}\sum_{i=1}^{D}\sum_{j=1}^{D} \tilde{g}^{-1/4} \frac{\partial}{\partial \xi_i} G_{i,j} \tilde{g}^{1/2} \frac{\partial}{\partial \xi_j} \tilde{g}^{-1/4}, \tag{2}$$

the fully rearranged form is practical in finite-basis representation (FBR) computations, *e.g.*, Ref. [17],

$$\hat{T}^V_{\text{frearr}} = -\frac{1}{2}\sum_{i=1}^{D}\sum_{j=1}^{D} G_{i,j} \frac{\partial}{\partial \xi_i}\frac{\partial}{\partial \xi_j} - \frac{1}{2}\sum_{j=1}^{D} B_j \frac{\partial}{\partial \xi_j} + U, \tag{3}$$

$$\text{with} \quad B_j = \sum_{i=1}^{D} \frac{\partial}{\partial \xi_i} G_{i,j}, \tag{4}$$

where $\mathbf{g} \in R^{(D+3)\times(D+3)}$ is the mass-weighted metric tensor, $\mathbf{G} = \mathbf{g}^{-1}$, $\tilde{g} = \det \mathbf{g}$, the extrapotential term,

$$U = \frac{1}{32}\sum_{k=1}^{D}\sum_{l=1}^{D}\left[\frac{G_{kl}}{\tilde{g}^2}\frac{\partial \tilde{g}}{\partial \xi_k}\frac{\partial \tilde{g}}{\partial \xi_l} + 4\frac{\partial}{\partial \xi_k}\left(\frac{G_{kl}}{\tilde{g}}\frac{\partial \tilde{g}}{\partial \xi_l}\right)\right], \tag{5}$$

and the volume element is $dV = d\xi_1 d\xi_2 \ldots d\xi_D$.

For atom-molecule complexes, we used a hybrid FBR-DVR scheme, including sin-cot-DVR [57] for the $c = \cos\theta$ degree of freedom, and the kinetic energy operator was written as [1, 21]

$$\hat{T}^V = -\frac{1}{2}\sum_{j=1}^{D}\frac{\partial}{\partial c} G_{c,j} \frac{\partial}{\partial \xi_j} - \frac{1}{2}\sum_{i=1,i\neq 2}^{D}\sum_{j=1}^{D} G_{i,j}\frac{\partial}{\partial \xi_i}\frac{\partial}{\partial \xi_j} - \frac{1}{2}\sum_{i=1}^{D} B_i \frac{\partial}{\partial \xi_i} + U, \tag{6}$$

$$B_i = \sum_{k=1,k\neq 2}^{D}\frac{\partial}{\partial \xi_k} G_{k,i}.$$

Since we aim to retain the full potential energy surface, we combine basis- [30, 31, 58, 59] and grid-pruning [18, 19, 20] techniques to make the 12-dimensional vibrational problem feasible for numerical computations. Efficient basis and grid pruning assumes good coordinates that are chosen to be spherical polar coordinates, $\xi_1 = R$, $\xi_2 = c = \cos\theta$, $\xi_3 = \varphi$, for the relative position of the fluoride around the methane fragment and normal coordinates, $\xi_{3+i} = q_i$ ($i = 1, 2, \ldots 9$) for the methane moiety [1]. (The same normal coordinate coefficients were used as in Ref. [1].)

The methane fragment is described with a non-direct product basis using the simple truncation condition for the basis indices (vibrational excitation 'quanta'), $n_{q_1} + \ldots + n_{q_9} \leq B = 4$, and the integrals were computed using a corresponding Smolyak grid [1, 21]. For the polar angle, $\theta$, we used 25 sin-cot-DVR points and functions. The $R$ and $\varphi$ degrees of freedom were described by 8 Morse tridiagonal basis functions (and 10 Gauss quadrature points) and 39 Fourier basis functions (and 42 trapezoidal quadrature points), respectively.

**B. Vibrational energies and splittings**

If tunneling among equivalent wells of the complex is prohibited, then the vibrational states can be classified according to the irreducible representations (irreps) of the $C_{3v}$ point group of the equilibrium structure. Due to the spread of the wave function over the four equivalent wells, the four-fold copies of the $C_{3v}$ irreps split up [1,60]

$$\Gamma(A_1^{C_{3v}}) = A_1 \oplus F_2,$$

$$\Gamma(A_2^{C_{3v}}) = A_2 \oplus F_1,$$

$$\Gamma(E^{C_{3v}}) = E \oplus F_1 \oplus F_2, \tag{7}$$

where $A_1$, $A_2$, $E$, $F_1$, and $F_2$ label irreps of the $T_d(M)$ molecular symmetry group.

The size of the splittings of the vibrational origins (**Table III**) is computed by 12D variational vibrational GENIUSH-Smolyak computations and the new PES developed in the present work. The tunneling splittings given in brackets in the table are estimated to be converged to the shown digits (*i.e.*, better than 0.1 cm$^{-1}$) using the largest basis and grid labelled with '$(B, C) = (4, 25)$'.

The $B = 4$ methane basis includes 715 intramolecular product functions and allows to converge the methane zero-point vibrational energy (ZPVE) to 0.66 cm$^{-1}$ [21]. In comparison with the $B = 4$ results, the $B = 3$ methane basis — which includes only 220 intramolecular methane functions and has a 2.51 cm$^{-1}$ error for the isolated methane description [21] — performs surprisingly well for the listed energies and splittings of CH$_4 \cdot$F$^-$ even at around 900

cm$^{-1}$ above the ZPVE, which is only ca. 400 cm$^{-1}$ lower than the lowest-energy intramolecular bending fundamental vibration of the methane fragment (at ca. 1300 cm$^{-1}$). Convergence using the $B$ = 2, 3, 4, . . . , 8 methane basis sets $n_{q_1} + \ldots + n_{q_9} \leq B$ has been tested for the isolated methane molecule in Ref. [21] (Table II). Regarding the tunneling splittings, splittings larger than 0.1 cm$^{-1}$ are clearly observable beyond 500 cm$^{-1}$, and several vibrational manifolds have splittings larger than 0.1 cm$^{-1}$ in the computed energy list, up to 903 cm$^{-1}$. The size of the splitting tends to be smaller for higher radial excitation (that correlates with the number of nodes, $n_R$, in **Table III**). If the radial excitation is lower, the vibrational energy has a larger fraction stored in the $\theta$, $\varphi$ intermolecular angular degrees of freedom and larger splittings are observed.

The electronic barrier height is 1269 cm$^{-1}$ on the present PES. The effective barrier height is smaller than this value, and it can be estimated by subtracting the 2D ($\theta$, $\varphi$) ZPVE that is on the order of 100 cm$^{-1}$ depending on the methane-ion separation (it is zero for an infinite methane-fluoride distance).

Beyond the top of the barrier, we expect to see the 'dissolution' of some tunneling manifolds, Eq. (7), in which the energy is stored in the intermolecular angular degrees of freedom. At the same time, if the energy is stored mostly in the methane-ion stretching and/or intramolecular (methane vibrational) modes, then the splittings are expected to be smaller—depending on the coupling strength of the intra-to-intermolecular dynamics. Computational study of these features will become accessible with further development of the vibrational methodology.

**Table III.** Vibrational band origins and tunneling splittings, $\tilde{v}$ and $\Delta$ in cm$^{-1}$, of CH$_4$·F$^-$ using the PES developed in this work and the GENIUSH-Smolyak algorithm [21]. The $\Delta$ (and $\Delta'$) splitting(s) give the separation of the neighboring states in a vibrational tunneling manifold, $\tilde{v}_1 = \tilde{v} + \Delta$ and $\tilde{v}_2 = \tilde{v}_1 + \Delta'$. No splitting values are given, if all vibrational energies in the manifold are within less than 0.1 cm$^{-1}$.

| # | $\Gamma$(MS) [a] | $n_R$ [b] | $\tilde{v}$ {$\Delta$} [this work] | | | | $\tilde{v}$ {$\Delta$} [21] |
|---|---|---|---|---|---|---|---|
| | | | (1, 25)[c] | (2, 25)[c] | (3, 25)[c] | (4, 25)[c] | (4, 25)[c] |
| 0–3 | $A_1 \oplus F_2$ | 0 | 9885.0 | 9858.8 | 9813.4 | 9806.7 | 9791.6 |
| 4–7 | $A_1 \oplus F_2$ | 1 | 190.8 | 192.1 | 192.4 | 192.6 | 193.6 |
| 8–15 | $E \oplus F_1 \oplus F_2$ | 0 | 264.9 | 262.0 | 263.8 | 262.9 | 267.6 |
| 16–19 | $A_1 \oplus F_2$ | 2 | 373.0 | 375.7 | 376.4 | 376.9 | 379.2 |
| 20–27 | $E \oplus F_1 \oplus F_2$ | 1 | 449.0 | 447.6 | 449.6 | 449.0 | 454.3 {0.1} |
| 28–31 | $A_1 \oplus F_2$ | 0 | 506.2 | 500.6 | 503.3 | 502.2 | 509.1 |
| 32–39 | $E \oplus F_1 \oplus F_2$ | 0 | 522.2 | 516.4 | 519.4 {0.0, 0.1} | 518.0 {0.0, 0.2} | 526.0 |
| 40–43 | $A_1 \oplus F_2$ | 3 | 547.3 | 551.8 | 553.1 | 554.1 | 557.5 |
| 44–51 | $E \oplus F_1 \oplus F_2$ | 2 | 624.3 | 624.5 | 626.8 | 626.6 | 632.7 |
| 52–55 | $A_1 \oplus F_2$ | 1 | 683.2 | 679.4 {0.1} | 682.3 | 681.4 {0.1} | 688.5 |
| 56–63 | $E \oplus F_1 \oplus F_2$ | 1 | 699.4 | 695.2 | 698.4 {0.0, 0.1} | 697.3 | 705.5 {0.0, 0.1} |
| 64–67 | $A_1 \oplus F_2$ | 4 | | 721.5 | 723.8 | 725.4 | 728.5 |
| 68–75 | $E \oplus F_1 \oplus F_2$ | 0 | | 728.0 {0.3, 0.1} | 731.4 {0.1, 0.4} | 730.1 {0.1, 0.3} | 738.1 {0.3, 0.4} |
| 76–83 | $E \oplus F_1 \oplus F_2$ | 0 | | 761.8 {0.1, 0.1} | 765.8 {0.1, 0.1} | 764.1 {0.1, 0.1} | |
| 84–91 | $E \oplus F_1 \oplus F_2$ | 3 | | 793.7 {0.1, 0.0} | 796.5 {0.0, 0.1} | 796.8 {0.1, 0.1} | |
| 92–95 | $A \oplus F$ [d] | 2 | | 849.0 {0.1} | 852.1 {0.1} | 851.6 {0.1} | |
| 96–103 | $E \oplus F_1 \oplus F_2$ | 2 | | 865.1 | 868.5 {0.1} | 867.8 | |
| 104–107 | $A \oplus F$ [d] | 5 | | 885.2 | 888.6 | 891.0 | |
| 108–115 | $E \oplus F_1 \oplus F_2$ | 1 | | 899.3 {0.1, 0.6} | 902.9 {0.1, 0.5} | 901.7 {0.2, 0.5} | |

[a] Symmetry assignment in the $T_d$(M) molecular symmetry group of the complex.

[b] Radial excitation (number of nodes) along the methane-ion separation, $R$, corresponding to the largest basis set, (4,25).

[c] ($B$, $C$) : The basis set corresponds to the $n_1 + \ldots + n_9 \leq B$ basis pruning condition for methane and $C$ gives the number of the sin-cot-DVR functions used for $\cos\theta$. The $R$ and $\varphi$ degrees of freedom were described by 8 Morse tridiagonal and 39 Fourier basis functions, respectively. The atomic masses used in the computations are $m$(H) = 1.007 825 032 23 u, $m$(C) = 12 u, and $m$(F) = 18.998 403 162 73 u [61].

[d] Symmetry assignment will be completed in future work when a larger number of states are computed.

## SUMMARY AND CONCLUSIONS

We have developed a new full-dimensional PES for the $CH_4 \cdot F^-$ anion complex by fitting high-level *ab initio* energies using the permutationally invariant polynomial approach. The new PES accurately describes the interaction between the methane and $F^-$ fragments at the complex region and along the dissociation coordinate. We have found that a physically motivated sampling of the configuration space involving randomly scattered $F^-$ around the randomly distorted $CH_4$ unit is not sufficient to obtain a high-quality PES for this strongly anisotropic complex. Therefore, for the first time in a spectroscopic application, we have used ROBOSURFER [44], which automatically develops the PES by selecting configurations from quasi-classical trajectories. Full-dimensional variational vibrational computations for the $CH_4$ fragment compared with experimental-quality results of isolated methane at different stages of the PES development have shown the outstanding accuracy of the final PES and the need of the machine-learning-type automatic sampling. The new PES opens the door for full-dimensional variational vibrational computations for the $CH_4 \cdot F^-$ anion complex. As a first step, in this study, we present vibrational band origins and tunneling splittings, utilizing the GENIUSH-Smolyak algorithm [21]. In the future, we plan to further test and/or improve our methodologies to access the intramolecular vibrational states of the methane fragment in this challenging complex system.

## ACKNOWLEDGEMENTS

The work of GC, DP, and VT at the University of Szeged was supported by the National Research, Development and Innovation Office−NKFIH, K-125317; the Ministry of Human Capacities, Hungary grant 20391-3/2018/FEKUSTRAT; Project no. TKP2021-NVA-19, provided by the Ministry of Innovation and Technology of Hungary from the National Research, Development and Innovation Fund, financed under the TKP2021-NVA funding scheme; and the Momentum (Lendület) Program of the Hungarian Academy of Sciences. EM and GA thank the financial support of the Swiss National Science Foundation (PROMYS Grant, No. IZ11Z0 166525).

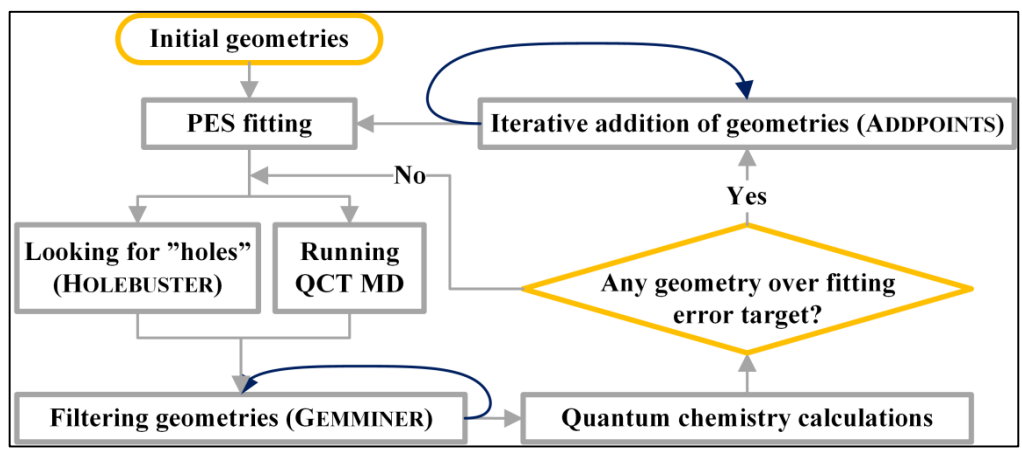

**Figure 1.** Schematic representation of the automated PES-development procedure of the ROBOSURFER program package. Taken from Ref. 44.

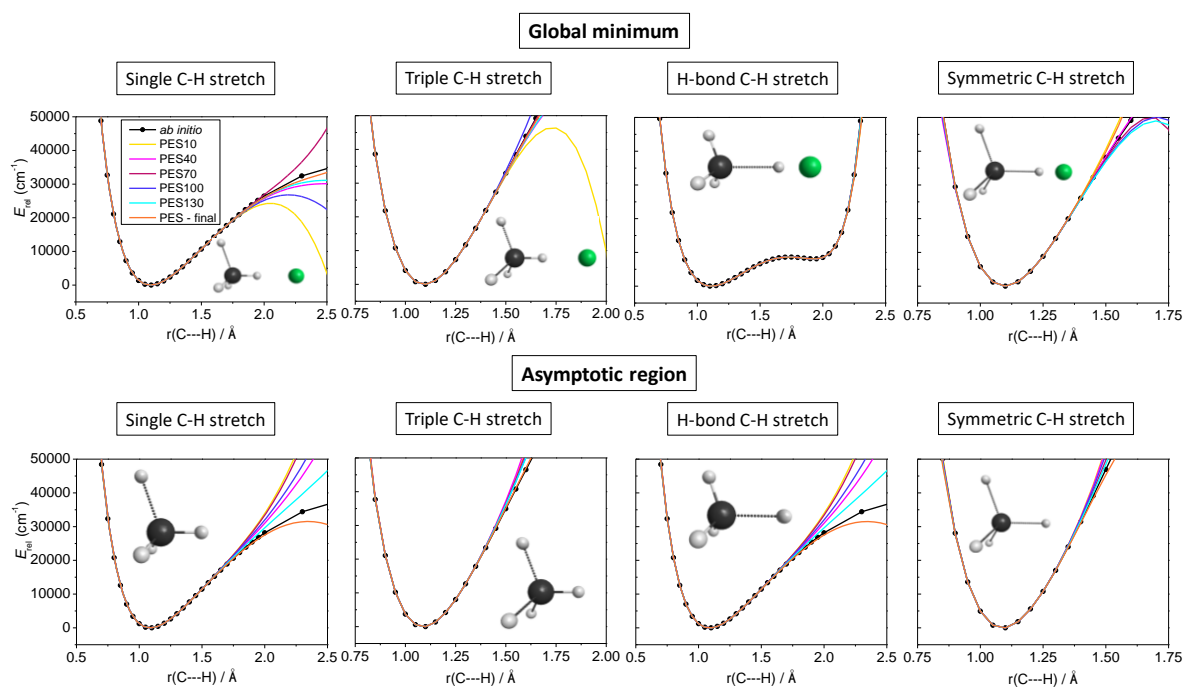

**Figure 2**. Evolution of one-dimensional potential energy cuts imitating various C-H stretching motions of methane in the global minimum structure or at the asymptotic region of the PES during the automated PES development using ROBOSURFER showing the 10$^{th}$, 40$^{th}$, 70$^{th}$, 100$^{th}$, and 130$^{th}$ iterations, compared to CCSD(T)-F12b/aug-cc-pVTZ *ab initio* energies. See text for details.

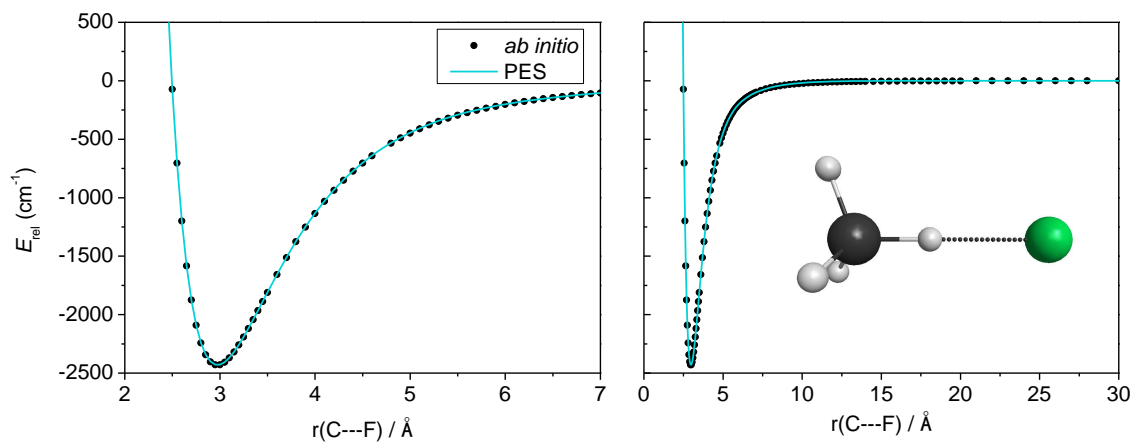

**Figure 3**. Potential energy cuts of the PES imitating the approach of F⁻ to methane along one C-H axis compared to the CCSD(T)-F12b/aug-cc-pVTZ *ab initio* energies.

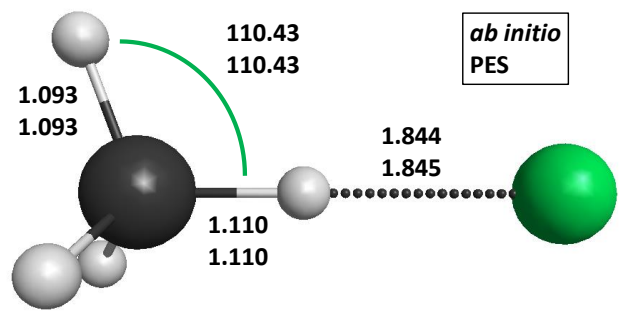

**Figure 4**. Equilibrium geometry of the global minimum of the $CH_4 \cdot F^-$ complex comparing structural parameters obtained at the CCSD(T)-F12b/aug-cc-pVTZ *ab initio* level of theory with those determined on the PES. Distances are in Å and the angle is in degree.